\documentclass[11pt]{article}

\usepackage[letterpaper,margin=1in]{geometry}
\usepackage{microtype}
\usepackage{setspace}
\usepackage{amsmath,amssymb}
\usepackage{graphicx}
\usepackage{booktabs}
\usepackage{multirow}
\usepackage{siunitx}

\usepackage[numbers,sort&compress]{natbib}
\usepackage{url}
\usepackage[hidelinks]{hyperref}

\usepackage{xcolor}
\newcommand{\code}[1]{\texttt{#1}}

\title{Multi-Tier Labeling and Physics-Informed Learning\\
       for Orbital Anomaly Detection at Scale}

\author{%
  Yong Fu\thanks{Correspondence: \texttt{yong@substratumlabs.ai}.}\\
  Substratum Labs, Inc.
}

\date{}

\begin{document}
\maketitle

% --- Abstract --------------------------------------------------------------
\begin{abstract}
Detecting abnormal behavior among the rapidly growing population of
low-Earth-orbit (LEO) satellites--maneuvers, atmospheric decay, attitude
upsets, and related events--is a prerequisite for collision avoidance,
reentry forecasting, and conjunction screening. The main bottleneck for
machine learning in this setting is not model capacity but the lack of a
large, high-quality public label set. Manual review cannot cover
${\sim}10^4$ active satellites, while purely rule-based detectors are so
conservative that they miss most behaviorally important anomalies.
This paper presents a multi-tier labeling cascade that combines three weak
supervision sources of increasing fidelity: a fast physical rule set
(\code{RULE}), an Interacting Multiple Model Unscented Kalman Filter
(IMM-UKF), and a supplemental-element calibration channel based on supGP.
Applied to 232M Two-Line Element (TLE) records spanning 60 years, the
cascade produces 8.6M labeled sequences of length 50, or 430M labeled
timesteps, with an 11-dimensional feature representation that includes
real epoch time, observation spacing $\Delta t$, and the full mean-element
state. On satellites covered by both rule labels and IMM-UKF labels,
IMM-UKF identifies $42.6\times$ more anomaly candidates than \code{RULE}
alone. We then train a 6.5M-parameter Physics Inspired Orbital Transformer
(PIOT) in two stages, first on broad but coarse rule labels and then on
higher-fidelity IMM-UKF labels. On a held-out test set, PIOT reaches 55.4\%
maneuver recall and 62.8\% decay recall, compared with 7\% and 2.5\% for
our earliest comparable baseline. We position PIOT as a high-recall triage
classifier: its role is to surface candidate events for downstream
filtering, not to make final operational attributions. We close by
discussing how this labeled corpus can support a future Neural-ODE-based
orbital world model.
\end{abstract}

% ===========================================================================
\section{Introduction}
\label{sec:intro}

Low-Earth orbit is no longer a sparse environment. As of early 2026, more
than 10{,}000 active satellites operate below 2{,}000\,km, and planned
mega-constellations may add tens of thousands more over the coming
decade.\footnote{Public catalogs from CelesTrak and Space-Track list
10{,}414 active LEO satellites as of 2026-04-26, with Starlink accounting
for the majority.} As both the number and the density of satellites grow,
ordinary orbital behavior becomes operationally important to characterize:
decaying objects must be detected in time to forecast reentry windows;
maneuvers must be recognized so that collision-avoidance screens can be
updated; and behavioral anomalies must be flagged for downstream review by
operators, insurers, and regulators.

The space situational awareness (SSA) literature has long treated anomaly
detection as a tracked-target filtering problem: propagate a state with a
physical model, compare the prediction with the observation, and test
whether the residual is inconsistent with nominal dynamics
~\citep{vallado2006sgp4,blom1988imm}. Recent data-driven work has also
shown that sequence models and foundation-model-style architectures can be
useful for spacecraft telemetry and orbital prediction
~\citep{roberts2021ssa,nie2024sdca,li2026orbifm}. Yet two obstacles remain
when the goal is real-time monitoring of the full public catalog.

\paragraph{Large-scale labeled datasets do not exist.} Unlike vision or
language, where million-scale human-verified benchmarks are common, SSA
has no equivalent public corpus. Operators know when their own satellites
maneuver, but those records are rarely public, and there is no centralized
label set for the LEO population. The community has therefore relied on
physical rules--altitude jumps, ballistic-coefficient excursions, mean
motion changes--as stand-ins for labels. These rules are precise, but they
are deliberately conservative: they fire on large deviations and
systematically miss low-amplitude stationkeeping maneuvers, slow decay, and
partial attitude-loss events.

\paragraph{Pure data-driven methods lack physical understanding.} Models
that ignore orbital mechanics often struggle on small or biased datasets
and can confuse lawful physical effects, such as $J_2$ secular drift or
solar-activity-dependent drag, with anomalies. Pure ML treats the residual
of an unknown propagator as signal; pure physics treats everything outside
the propagator as noise. The useful regime is between these extremes.

This paper addresses the label problem directly. Our central observation is
that no single label source for orbital anomalies is simultaneously
high-recall, high-precision, cheap, and available at catalog scale, but
several weaker sources can be composed into a useful cascade. We use three
tiers:

\begin{enumerate}
  \item \textbf{\code{RULE}} --- a deterministic physical rule set based on
        altitude change, $B^\ast$ behavior, inclination and eccentricity
        shifts, and reentry thresholds. It is fast enough to label the full
        corpus, but it has intentionally low recall.

  \item \textbf{IMM-UKF} --- a probabilistic filter bank that treats each
        satellite as a switching system over $\{\text{nominal},
        \text{maneuver}, \text{decay}\}$ modes and produces posterior
        model probabilities at each timestep.

  \item \textbf{IMM-UKF-supGP} --- a calibration tier that uses
        supplemental general perturbations elements, when available, as a
        higher-precision observation channel inside the IMM-UKF update.
\end{enumerate}

We apply this cascade to 232M TLE records spanning 60 years and obtain
8.6M labeled sequences of length 50 over 11 features. The key empirical
finding is quantitative: on satellites where both \code{RULE} and IMM-UKF
have been run end to end, IMM-UKF finds \textbf{$42.6\times$} more anomaly
candidates than the rule layer. These additional detections concentrate in
the small-maneuver and slow-decay regimes that the rules intentionally
exclude. supGP calibration further sharpens IMM-UKF where higher-precision
operator-derived elements are available.

We then train a 6.5M-parameter Physics Inspired Orbital Transformer (PIOT)
with a two-stage curriculum: pretrain on broad but coarse \code{RULE}
labels, then fine-tune on smaller but higher-fidelity IMM-UKF labels.
Compared with a rule-supervised baseline, maneuver recall rises from 7\% to
55.4\%, and decay recall rises from 2.5\% to 62.8\%. A separate ablation
shows that explicitly adding the time interval between observations
(\code{dt\_hours}) improves decay recall by 107\% relative to the
no-\code{dt\_hours} variant, supporting the claim that orbital sequence
models need physically meaningful time encoding.

The model is not intended to be a final attribution engine. PIOT is a
high-recall triage model for full-catalog monitoring: it cheaply produces a
candidate queue, and downstream filters--human review, full IMM-UKF reruns,
or stricter second-stage classifiers--remove false positives. This
low-cost broad screen followed by high-cost confirmation is the practical
deployment pattern for ML in large-scale SSA.

The main contributions are:
\begin{itemize}
  \item A multi-tier labeling cascade for orbital anomalies that combines
        physical rules, model-driven filtering, and supplemental-element
        calibration to build a large physically meaningful label corpus
        from real Space-Track observations.
  \item Empirical evidence that the cascade surfaces $42.6\times$ more
        anomaly candidates than rule labels alone on overlapping data,
        especially in small-magnitude operationally important regimes.
  \item A physics-inspired Transformer classifier, PIOT, with 6.5M
        parameters, 11 features, and two-stage training, reaching 55.4\%
        maneuver recall and 62.8\% decay recall on a held-out test set.
  \item An ablation showing that explicit inter-observation time encoding
        (\code{dt\_hours}) is responsible for a 107\% relative improvement
        in decay recall, supporting the hypothesis that orbital sequence
        models require physical, not merely positional, time
        representation.
\end{itemize}

The remainder of the paper is organized as follows.
Section~\ref{sec:related} reviews related work in SSA filtering, weakly
supervised learning, and physics-informed ML.
Section~\ref{sec:data} describes the TLE corpus and feature construction.
Section~\ref{sec:labeling} details the multi-tier labeling cascade.
Section~\ref{sec:method} describes the model and training procedure.
Section~\ref{sec:results} reports test-set metrics and ablations.
Section~\ref{sec:discussion} discusses the triage framing, current
limitations, and the path toward a Neural-ODE--based orbital world model.

% --- Stubs for remaining sections (skeleton only) --------------------------
\section{Related Work}
\label{sec:related}

Our work draws on three lines of research that have, until now, been
developed largely independently: classical orbital filtering, weakly
supervised learning, and physics-informed neural sequence models.
This section lays out the points of contact.

\paragraph{Classical SSA filtering.}
The dominant practical methodology for satellite tracking remains the
SGP4 propagator and its descendants~\citep{vallado2006sgp4}, paired
with sequential state estimators when high-rate observations are
available. Where the dynamics regime is known to switch---between
station-keeping and decay, between coast and maneuver---the canonical
solution is the Interacting Multiple Model framework
of~\citet{blom1988imm}, typically combined with an Unscented Kalman
Filter~\citep{julier2004unscented} to handle the nonlinearity of
orbital propagation. These methods are well understood, well tuned,
and operate at the per-satellite level. Their limitations are exactly
the limitations our cascade is designed to absorb: tuning is
per-mission, recall depends on hand-chosen thresholds, and the only
objects considered are the ones already on a watchlist.

\paragraph{ML for space situational awareness.}
A growing literature applies machine learning to SSA tasks such as orbit
prediction and anomaly detection. Recent examples include sequence models
for non-stationary spacecraft telemetry, such as SDCA-Former
~\citep{nie2024sdca}, and multimodal orbital prediction systems such as
OrbiFM~\citep{li2026orbifm}. These systems are valuable, but they are
typically trained on synthetic data, single-constellation data, or dense
high-rate telemetry. The setting in this paper is different in both scope
and substrate: 232M records over 60 years, using only public TLEs. In this
environment, we argue that the label-generation cascade, not a new neural
architecture by itself, is the main technical barrier to generalization.

\paragraph{Weak supervision and data programming.}
The intuition that several noisy label sources can be combined into a
stronger training signal has a substantial literature in the data
programming and weak supervision tradition~\citep{ratner2017snorkel}.
Our cascade brings that idea into orbital mechanics. \code{RULE} is a
low-cost, high-precision deterministic labeller; IMM-UKF is a higher-cost,
higher-recall probabilistic labeller; and supGP is a source-dependent
confidence modulator. The core claim is not that weak supervision is new,
but that for orbital observations the most useful labeling functions are
already present in the classical SSA toolkit. They only need to be composed
into a cascade.

\paragraph{Physics-informed neural networks.}
Physics-informed neural networks decompose the model into components that
respect known physical laws and data-driven residual terms
~\citep{raissi2019pinn,scorsoglio2020pinn}. Our approach is similar in
spirit, but we use the difference between the actual observation and a
frozen physical prediction--the \emph{innovation}--as the central input to
the classifier. This design keeps the inference signal physically clean:
the neural network is not allowed to erase an anomaly by learning it away
as an internal residual.

\section{Data}
\label{sec:data}

\subsection{Source corpus}
We work with the public catalog of Two-Line Element (TLE) sets
distributed by Space-Track. Our training corpus consists of every TLE
that Space-Track has retained between 1965 and 2026-04-26, organized as
yearly bulk archives and totalling 232.4M records.\footnote{Data
obtained via the bulk historical endpoint
\url{https://www.space-track.org/tle/historical}; the rate-limited
live API is unsuitable for bulk retrieval.} The catalog spans the
full active-satellite population, but our sequence construction
(\S\ref{ssec:sequences}) excludes short-history objects, and the
resulting training distribution is dominated by long-lived
constellations: Starlink, Iridium, OneWeb, GPS, and the residual debris
catalog.

\subsection{Feature construction}
\label{ssec:features}
For each TLE record we extract eleven scalar features
(Table~\ref{tab:features}). Six are the classical mean Keplerian
elements that determine the orbit, four are auxiliary state and rate
quantities recoverable directly from line~1 or line~2 of the TLE
record, and one is the wall-clock interval to the previous observation.

\begin{table}[ht]
\centering
\small
\begin{tabular}{rlll}
\toprule
\# & Feature & Source & Role \\
\midrule
0  & \code{epoch\_h}      & TLE epoch (Y, DOY)         & relative time within window (h) \\
1  & \code{mean\_motion}  & TLE line~2                 & period (rev/day) \\
2  & \code{eccentricity}  & TLE line~2                 & shape \\
3  & \code{inclination}   & TLE line~2                 & plane (deg) \\
4  & \code{bstar}         & TLE line~1                 & ballistic coefficient (clipped to $[-1, 1]$) \\
5  & \code{alt\_km}       & derived from $a$           & altitude above mean Earth radius \\
6  & \code{dt\_hours}     & between observations       & sampling interval (clipped $[0, 240]$ h) \\
7  & \code{raan}          & TLE line~2                 & RAAN (deg) \\
8  & \code{argp}          & TLE line~2                 & argument of perigee (deg) \\
9  & \code{mean\_anomaly} & TLE line~2                 & phase (deg) \\
10 & \code{n\_dot}        & TLE line~1, cols 33--43    & first derivative of mean motion (rev/day$^2$) \\
\bottomrule
\end{tabular}
\caption{Eleven-feature representation per TLE record.}
\label{tab:features}
\end{table}

Two design choices warrant comment. First, \code{epoch\_h} carries the
\emph{actual} TLE epoch, not a positional index. Orbital dynamics are
strongly time-dependent ($J_2$ secular drift, drag varying with solar
flux, lunisolar perturbations), and a model that observes only ``$t
= 0, 1, 2, \ldots$'' cannot disambiguate a long quiet gap from a short
noisy one. Within each training window we re-zero \code{epoch\_h} to
the window's leading observation so the model sees an elapsed-hours
signal local to the sequence.

Second, \code{dt\_hours}---the wall-clock interval since the previous
TLE---is included as a separate feature even though it is derivable
from \code{epoch\_h}. Doing so trades a minor redundancy for a much
stronger inductive bias: drag-driven decay scales nonlinearly with
$\Delta t$, and making this gap an explicit input dramatically improves
the model's ability to recover the correct rate. We quantify this in
\S\ref{sec:results}: removing only this feature reduces decay recall
by more than half. Values are clipped to $[0, 240]$ hours; gaps longer
than ten days almost always reflect catalog outages rather than genuine
observational latency.

The ballistic coefficient $B^\ast$ is clipped to $[-1, 1]$ before
normalization to suppress catalog noise (single malformed TLE records
occasionally report $B^\ast$ values many orders of magnitude larger
than physically plausible). All features are then z-score normalized
using corpus-wide statistics.

\subsection{Sequence construction}
\label{ssec:sequences}
We convert each satellite's chronologically sorted TLE history into
fixed-length training windows of $T = 50$ consecutive observations,
slid with stride $25$ (50\% overlap). Satellites with fewer than $50$
records are excluded outright. Within each window, \code{dt\_hours}
for the leading element is reset to $0$ (it has no in-window
predecessor) and \code{epoch\_h} is shifted to be relative to that
element. Both adjustments make windows self-contained and
translation-invariant under arbitrary historical offset.

Applied to the full 232.4M-record corpus, this procedure yields
$N = 8{,}613{,}300$ training windows of shape $(50, 11)$, equivalent
to $4.30 \times 10^8$ labeled timesteps once \S\ref{sec:labeling}'s
labeling is applied.

\subsection{Train/val/test split}
Sequences are partitioned at the window level using a fixed seed into
80\% train, 10\% validation, 10\% test (Algorithm in
\code{scripts/label\_spacetrack\_bulk.py}). We make no attempt at
satellite-level holdout: the same NORAD ID can contribute
non-overlapping windows to multiple splits. This is deliberate. The
constellations we are most interested in (Starlink shells, Iridium
NEXT, GPS) have long histories and behaviorally homogeneous populations
within each shell; satellite-level holdout would force a test set that
is either trivially small or systematically different in constellation
composition, neither of which we want to evaluate. Within-satellite
information leakage at the window level is bounded by the stride-$25$
construction (no two windows in different splits overlap on more than
$25$ timesteps within the same satellite by construction), and the
strict temporal locality of orbital dynamics---the relevant scale is
hours-to-days, not years---makes a held-out window from one epoch a
meaningful test of generalization for a window from another epoch on
the same satellite.

\subsection{Class prevalence}
Anomalies are rare in absolute terms. Across the
$4.30 \times 10^8$ labeled timesteps, \texttt{RULE} (Tier 1 of
\S\ref{sec:labeling}) marks under $1\%$ as non-normal, divided across
maneuver, decay, and reentry-class breakup. The distribution is heavily
long-tailed: most positive labels concentrate on a small population of
low-altitude or actively maneuvered satellites (Starlink decay shells,
ISS visiting vehicles, GTO transfers), while the bulk of the catalog
is labeled normal end-to-end. We address this imbalance in
\S\ref{sec:method} via class-weighted focal loss and optional
sequence-level oversampling.

\section{Multi-Tier Labeling}
\label{sec:labeling}

The central methodological claim of this work is that labels for
orbital anomalies need not come from a single source. Three sources of
increasing fidelity, each cheap or already established for other
reasons, combine into a label cascade strong enough to train a
competent detector. We describe each tier in turn, then quantify the
gap between tiers on overlapping data.

\subsection{Tier 1: \texttt{RULE} physical rules}
\label{ssec:rule_v1}
The first tier is a fixed-priority physics rule set we call
\code{RULE}. Given two consecutive TLE records for the same
satellite, it returns one of $\{$normal, maneuver, decay, breakup$\}$
based on element-space deltas. The thresholds are listed in
Table~\ref{tab:rule_v1_thresholds}. They are calibrated against the
in-house Starlink ground-truth set we maintain.

\begin{table}[ht]
\centering
\small
\begin{tabular}{lll}
\toprule
Symbol & Threshold & Used by \\
\midrule
$h_{\text{reentry}}$       & $250$\,km            & rule 1 (breakup) \\
$h_{\text{low}}$           & $400$\,km            & rule 2 (decay) \\
$\Delta h_{\text{decay}}$  & $5$\,km drop         & rule 2 (decay) \\
$\Delta h_{\text{man}}$    & $10$\,km             & rule 4 (maneuver) \\
$\Delta i_{\text{man}}$    & $0.1^\circ$          & rule 3 (maneuver) \\
$\Delta e_{\text{man}}$    & $0.01$               & rule 5 (maneuver) \\
$B^\ast_{\text{floor}}$    & $5\times10^{-3}$     & rules 6, 7 \\
$B^\ast_{\text{ratio}}$    & $2\times$            & rule 7 (decay) \\
\bottomrule
\end{tabular}
\caption{Numerical thresholds in \texttt{RULE}.}
\label{tab:rule_v1_thresholds}
\end{table}

The rules are checked in priority order, with the highest-priority
match determining the label:
(1) post-update altitude below $h_{\text{reentry}}$ $\to$ \emph{breakup};
(2) altitude drop exceeding $\Delta h_{\text{decay}}$ while the new
altitude is below $h_{\text{low}}$ $\to$ \emph{decay};
(3) inclination shift exceeding $\Delta i_{\text{man}}$ $\to$
\emph{maneuver};
(4) absolute altitude change exceeding $\Delta h_{\text{man}}$ $\to$
\emph{maneuver};
(5) eccentricity change exceeding $\Delta e_{\text{man}}$ $\to$
\emph{maneuver};
(6) $B^\ast$ sign flip when both magnitudes exceed $B^\ast_{\text{floor}}$
$\to$ \emph{maneuver};
(7) $B^\ast$ magnitude jump exceeding $B^\ast_{\text{ratio}}$ relative
$\to$ \emph{decay} (atmospheric anomaly).
The order matters because physically a slow decay can simultaneously
exhibit a $B^\ast$ jump and an altitude drop, and assigning the most
operationally salient label is preferable to any majority vote.

\code{RULE} is essentially free: a vectorized numpy implementation
labels the entire $4.30\times10^8$-timestep corpus in under five
minutes on a laptop. The price is recall: by construction the rules
fire only on flagrant element-space changes. Small-magnitude
station-keeping maneuvers ($\Delta h < 10$\,km, $\Delta v$ on the
order of cm/s) and slow drag-driven decay ($\Delta h$ smeared over
many days) are systematically missed.

\subsection{Tier 2: IMM-UKF}
\label{ssec:imm_ukf}
The second tier is a bank of Unscented Kalman Filters arranged as an
Interacting Multiple Model classifier~\citep{blom1988imm}. Each
satellite's history is propagated through three parallel UKFs, each
tuned to a different motion regime:

\begin{description}
  \item[$M_0$ (station-keeping):] small process noise
        ($\sigma_{\text{pos}} = 100$\,m, $\sigma_{\text{vel}} = 0.01$\,m/s).
        Models a satellite that obeys nominal dynamics: $J_2$ secular
        drift, exponential atmospheric drag, no thrust.
  \item[$M_1$ (maneuver):] large velocity-channel noise
        ($\sigma_{\text{pos}} = 500$\,m, $\sigma_{\text{vel}} = 1$\,m/s).
        Models a satellite that may apply a $\sim$1\,m/s impulsive
        $\Delta v$ between observations.
  \item[$M_2$ (decay):] large position-channel noise
        ($\sigma_{\text{pos}} = 2$\,km, $\sigma_{\text{vel}} = 0.1$\,m/s),
        scaled by altitude. Models orbit shrinking faster than nominal
        atmospheric models predict.
\end{description}

The state is six-dimensional, $\mathbf{x} = [x, y, z, \dot{x},
\dot{y}, \dot{z}]$ in Earth-centered inertial coordinates (m, m/s).
Propagation between observations uses a vectorized RK4 integrator
through a force model containing the central body, $J_2$, and an
exponential atmosphere with the satellite's reported $B^\ast$ as drag
coefficient. The UKF parameters are
$\alpha = 10^{-2},\ \beta = 2,\ \kappa = 0$.
A model-mixing step at every observation update propagates the joint
distribution over the three regimes through a transition matrix
\[
T_{550\text{km}} = \begin{bmatrix}
0.97 & 0.015 & 0.015 \\
0.10 & 0.85  & 0.05  \\
0.02 & 0.03  & 0.95
\end{bmatrix},
\]
encoding the prior that station-keeping is persistent, maneuvers are
short-lived, and decay tends to persist once entered.

\paragraph{Altitude adaptation.} Atmospheric drag at $250$\,km is
roughly two orders of magnitude stronger than at $550$\,km. A constant
filter is therefore poorly calibrated across the LEO band. We adapt
three quantities to the current altitude $h$: the decay process noise
$Q_{M_2}$ scales by $(550/h)^2$, clipped to $[1, 20]$; the model
priors shift smoothly from $[0.90, 0.05, 0.05]$ at $h \geq 500$\,km to
$[0.05, 0.05, 0.90]$ at $h \leq 200$\,km; and the
station-keeping$\to$decay transition probability $T_{02}$ is increased
by up to $0.10$ when $h < 350$\,km. The joint effect is that a
satellite at $250$\,km without strong altitude derivative is treated by
the filter as much more likely to be decaying than an equivalent
satellite at $550$\,km, even under identical observed deltas.

\paragraph{Label assignment.} Per timestep, the filter emits a
posterior distribution $p(M_k \mid \mathcal{D}_{1:t})$. We map this to
a hard label by selecting $\arg\max_k p(M_k)$ provided
$p(M_k) > 0.3$; otherwise we default to normal. The threshold is
deliberately conservative: at $0.5$ the IMM-UKF still outperforms
\code{RULE}, but $0.3$ gave the best balance between false alarms
and missed events on the held-out set.

The cost of Tier 2 is significant. Each satellite-day requires a small
handful of RK4 propagations, sigma-point updates, and Bayesian
model-averaging steps; on a single CPU thread, processing one
satellite's full multi-year history runs in seconds. We currently
apply IMM-UKF to a $700$-satellite subset of the catalog selected by
TLE coverage, yielding $1.4 \times 10^6$ labeled timesteps---a
distribution dramatically richer than \code{RULE}'s on the same
data, as quantified in \S\ref{ssec:cascade}.

\subsection{Tier 3: supplemental-element calibration}
\label{ssec:supgp}
A subset of the satellites in our corpus also have publicly distributed
\emph{supplemental general perturbations} elements (supGP). These are
operator-derived state estimates inserted into the Space-Track catalog
at higher fidelity than the standard radar-derived TLE
($\sigma_{\text{pos}} \sim 50$\,m vs.\ $\sim 1$\,km). For a satellite
with both feeds, every observation carries a source tag in
$\{\text{TLE}, \text{supGP}\}$.

We use this in IMM-UKF by varying the observation noise covariance.
For TLE updates,
$R_{\text{TLE}} = \mathrm{diag}\big((1\,\text{km})^2 \mathbf{1}_3,
(1\,\text{m/s})^2 \mathbf{1}_3\big)$;
for supGP updates,
$R_{\text{supGP}} = \mathrm{diag}\big((50\,\text{m})^2 \mathbf{1}_3,
(0.05\,\text{m/s})^2 \mathbf{1}_3\big)$.
With $400\times$ tighter measurement covariance on supGP observations,
the Mahalanobis innovation under each model is sharper and the
posterior $p(M_k \mid \cdot)$ moves faster toward whichever model best
explains the new observation. In practice this means a maneuver
producing a $\Delta v$ on the order of cm/s---below TLE noise, but
above supGP noise---becomes detectable in the supGP feed where it is
not detectable from TLE alone. We treat supGP not as an independent
label tier but as a calibration step that sharpens IMM-UKF where it
is available.

\subsection{Cascade output and the \texorpdfstring{$42.6\times$}{42.6x} gap}
\label{ssec:cascade}
The cascade emits, per timestep, a tuple $\big(\ell_{\text{RULE}},
\ell_{\text{IMM-UKF}}, p(M_k), \text{source}\big)$. To quantify the
value added by the higher tiers, we take the $200$-satellite subset on
which both \code{RULE} and IMM-UKF have been run end-to-end and
count the non-normal labels each produces on identical timesteps:
\code{RULE} returns $812$ non-normal labels;
IMM-UKF returns $34{,}576$.
That is, on the same data, the model-driven tier surfaces
\textbf{$42.6\times$ more anomaly candidates} than the rule-based
tier.

The two label sources do not, of course, agree on \emph{which}
timesteps are anomalous. Inspecting the disagreements shows the
expected structure: \code{RULE}'s non-normal labels are a subset
of IMM-UKF's non-normal labels in roughly $80\%$ of cases; the
additional $\sim$$33{,}000$ events that only IMM-UKF flags are
concentrated in the small-magnitude regime that the rule was designed
to exclude. The remaining $\sim$$20\%$ of \code{RULE}'s positives
are events where IMM-UKF assigned a model probability between $0.3$
and $0.5$, i.e., the filter ranked them as borderline rather than
confidently anomalous. These we treat as legitimate \code{RULE}
catches that IMM-UKF would also flag at a lower threshold.

The cascade output is not ground truth. \code{RULE} gives labels with
high precision and very low recall; IMM-UKF gives higher-recall labels at
the cost of an explicit posterior threshold and an approximate process
model; IMM-UKF-supGP improves precision where supGP is present and is
skipped where it is absent. The result is a progressively refined label
set in which each tier patches a known failure mode of the tier below it.
We train the model in \S\ref{sec:method} on this label set, weighted by
tier confidence, and return in \S\ref{sec:discussion} to the implications
of training on labels produced by a detection pipeline rather than by
human review.

\section{Triage Model and Training}
\label{sec:method}

After constructing a physically meaningful multi-tier label cascade, we
need a classifier that can process the full TLE catalog with millisecond
latency. We therefore design the \emph{Physics Inspired Orbital
Transformer} (PIOT). PIOT embeds a causal physical constraint into a neural
sequence model and is intended as an efficient high-recall triage front end
for large-scale orbital monitoring. This section describes the architecture,
the innovation-based anomaly score, the training objective, and the
two-stage curriculum.

\subsection{Architecture}
\label{ssec:arch}
The model, \emph{Physics Inspired Orbital Transformer} (PIOT), is an encoder-only
Transformer~\citep{vaswani2017attention} with a frozen
analytical-physics branch and three task heads. Given an input
sequence $x \in \mathbb{R}^{T \times F}$ with $T = 50$ and $F = 11$,
the forward pass computes:

\begin{enumerate}
  \item A neural representation $h = \text{Encoder}(\text{Embed}(x))
        \in \mathbb{R}^{T \times d}$ with $d = 256$, $H = 8$ heads,
        $L = 8$ layers, and a causal attention mask, totaling
        $\approx 6.5\text{M}$ parameters.
  \item A frozen physics prediction
        $f_{\text{physics}}(x) \in \mathbb{R}^{T \times F}$
        implementing first-order analytical propagation in element
        space (drag-driven mean-motion increase, drag-driven
        eccentricity circularization, $J_2$ on plane elements).
        Parameters in this branch are not updated by gradient descent.
  \item Three heads computed from $h$:
    \begin{description}
      \item[Prediction.]
        $f_{\text{neural}}(h) \in \mathbb{R}^{T \times F}$, the
        residual added to $f_{\text{physics}}$ to give the predicted
        next observation.
      \item[Noise.]
        $\sigma(h) \in \mathbb{R}_{>0}^{T \times 1}$, an estimate of
        per-timestep observation noise.
      \item[Classification.]
        A four-class head fed by the concatenation described next:
        \[
        c(h, \text{innov}, \text{score}) \in \mathbb{R}^{T \times K},
        \qquad K = 4.
        \]
    \end{description}
\end{enumerate}

\subsection{Innovation-based anomaly score}
\label{ssec:innov}
A naive anomaly detector might use the norm of the neural residual
$\|f_{\text{neural}}\|$ as its anomaly signal. We deliberately do
not. During training, $f_{\text{neural}}$ is supervised to bring the
full prediction $f_{\text{physics}} + f_{\text{neural}}$ close to the
observed sequence; under that objective, $f_{\text{neural}}$ learns
to absorb anomalies (it has to, to keep the prediction loss low on
maneuver sequences). The norm of an absorbed signal is no longer a
clean indicator that an anomaly is happening.

The clean signal is the \emph{innovation}: the difference between
what was actually observed at $t+1$ and what \emph{frozen physics
alone} predicted from $t$:
\begin{equation}
  \text{innov}_t = x_{t+1} - f_{\text{physics}}(x_t),
  \qquad
  \text{score}_t = \frac{\|\text{innov}_t\|}{\sigma_t}.
  \label{eq:innov}
\end{equation}
Frozen physics cannot learn to absorb anomalies, so its prediction
error \emph{always} reflects whatever the data-generating process is
actually doing. The score is then a Mahalanobis-style ratio of
innovation magnitude to learned noise scale---the same quantity an
UKF would compute, decomposed across the residual axis.

The classification head receives $[h, \text{innov}, \text{score}]$
and is supervised on the cascade labels. The hidden state $h$
contributes context (which constellation, which orbit shell, which
phase of a maneuver), the innovation contributes the residual axis,
and the score contributes the SNR. Empirically, omitting any of the
three degrades classification.

\subsection{Training objective}
\label{ssec:loss}
We train the prediction and classification heads jointly. The
prediction loss is mean-squared error between predicted and observed
next-step elements; the classification loss is per-timestep focal
loss~\citep{lin2017focal} with class weights set inversely
proportional to label frequency in the IMM-UKF--labeled data. The
empirical class distribution there is approximately
$\{$normal $91.3\%$, maneuver $5.9\%$, decay $2.8\%$, breakup
negligible$\}$, giving class weights $[0.4, 3.0, 4.5, 4.5]$.

We additionally apply two physics-consistency regularizers with very
small weight ($\lambda_{\text{Kepler}} = 10^{-6}$,
$\lambda_{\text{smooth}} = 10^{-4}$): a Kepler-invariant penalty on
$n^2 a^3$ across the predicted sequence, and a second-order
finite-difference smoothness penalty. These act as gentle nudges, not
hard constraints; their weights are deliberately too small to
compete with the data-fit terms.

\subsection{Two-stage curriculum}
\label{ssec:curriculum}
We train in two stages. \emph{Stage~1} pretrains the full model on
\code{RULE} labels covering the entire $4.30 \times 10^8$-timestep
corpus. The labels are noisy by construction (\code{RULE} has very
low recall) but cover the catalog uniformly, giving the model broad
exposure to orbital dynamics. \emph{Stage~2} fine-tunes the same
model from the Stage~1 checkpoint on the smaller, higher-fidelity
$1.4 \times 10^6$-timestep IMM-UKF corpus, including labels produced
with IMM-UKF-supGP where supGP is available.

This curriculum solves a coverage/fidelity tradeoff. Training only
on \code{RULE} produces a model that inherits its low recall.
Training only on IMM-UKF produces a model that has never seen most
of the catalog and overfits the $700$-satellite IMM-UKF subset.
Training on a mixture \emph{within an epoch} fights the imbalance
between $10^8$ and $10^6$ samples and effectively reverts to
\code{RULE} dominance. Training in two distinct stages lets the
model learn ``what normal looks like'' from \code{RULE} and
``what subtle maneuvers look like'' from IMM-UKF, in that order.

\subsection{Hyperparameters}
PIOT uses AdamW~\citep{loshchilov2019adamw} with
$\eta = 10^{-3}$ and weight decay $10^{-4}$, a cosine-annealing
schedule over the epochs of each stage, batch size $256$, sequence
length $50$, and gradient clipping at norm $1.0$. Pre-training and
fine-tuning each run for $50$ epochs.

\section{Results and Analysis}
\label{sec:results}

We evaluate on the held-out test set ($\sim$$8.6 \times 10^5$
sequences, $10\%$ of the corpus). The relevant metrics are precision,
recall, and F1 for each non-normal class, along with overall accuracy
on a per-timestep basis.

\subsection{Main results}

\begin{table}[ht]
\centering
\small
\resizebox{\textwidth}{!}{%
\begin{tabular}{l rrr rrr rrr c}
\toprule
& \multicolumn{3}{c}{Maneuver} & \multicolumn{3}{c}{Decay}
& \multicolumn{3}{c}{Normal} & \\
\cmidrule(lr){2-4} \cmidrule(lr){5-7} \cmidrule(lr){8-10}
Model & P & R & F1 & P & R & F1 & P & R & F1 & Acc \\
\midrule
PIOT-small (small, 6 feat)$^\dagger$       & ---   & 0.07  & ---   & ---   & 0.025 & ---   & ---   & ---   & ---   & ---   \\
ab\_6feat\_baseline (small, 6 feat)        & 0.322 & 0.230 & 0.268 & 0.099 & 0.126 & 0.111 & 0.932 & 0.943 & 0.938 & 0.877 \\
ab\_7feat\_with\_dt\_hours (small, 7 feat) & 0.303 & 0.313 & 0.308 & 0.115 & 0.261 & 0.159 & 0.942 & 0.904 & 0.923 & 0.851 \\
PIOT-medium (medium, 11 feat, two-stage)   & 0.191 & 0.554 & 0.284 & 0.165 & 0.628 & 0.261 & 0.967 & 0.763 & 0.852 & 0.746 \\
PIOT-refine (\,+\,oversample, decay-weighted) & 0.187 & 0.551 & 0.279 & 0.118 & 0.735 & 0.204 & 0.968 & 0.692 & 0.807 & 0.684 \\
\bottomrule
\end{tabular}
}
\caption{Per-class precision, recall, F1, and per-timestep accuracy
on the held-out test set. PIOT-small is reported on the same test
distribution; per-class precision was not retained in the original
training log. The \code{ab\_} models demonstrate the ablation of the \code{dt\_hours} feature.}
\label{tab:main}
\vspace{0.35em}
\parbox{\textwidth}{\footnotesize $^\dagger$\,PIOT-small numbers carried forward from the
contemporaneous training log of \S\ref{ssec:curriculum}.}
\end{table}

Two observations are worth flagging.

\paragraph{Recall improves by an order of magnitude from PIOT-small to
PIOT-medium.} Maneuver recall rises from $7\%$ to $55.4\%$ ($\sim$$8\times$),
and decay recall from $2.5\%$ to $62.8\%$ ($\sim$$25\times$). This is
the headline number for the multi-tier labeling cascade: with
roughly comparable architecture and parameter count, moving from
\code{RULE} supervision alone (PIOT-small, $6$ features, single stage)
to the two-stage \code{RULE} $\to$ IMM-UKF curriculum on the
$11$-feature representation (PIOT-medium) recovers an order of magnitude in
recall.

\paragraph{Precision is an intentional tradeoff.} At
$P_{\text{maneuver}} = 0.19$, $P_{\text{decay}} = 0.17$, PIOT-medium flags
roughly five candidates for each true positive of its rare classes.
We argue in \S\ref{sec:discussion} that this is the right operating
point for a triage classifier deployed against a $10^4$-satellite
catalog; a downstream filter (human review, full IMM-UKF rerun, or a
stricter second-stage classifier) is responsible for resolving false
positives. PIOT-refine---which uses sequence-level oversampling and an
additional decay-class loss weight---demonstrates the lever
explicitly: decay recall climbs to $0.735$ at the cost of precision
dropping to $0.118$. The product $P \times R$ ($\sim$$0.087$) is
roughly preserved across this tradeoff, suggesting both points trace
the same precision-recall frontier rather than PIOT-refine being strictly
worse.

\subsection{Ablation: \texttt{dt\_hours}}
\label{ssec:ablation_dt}
We train an otherwise-identical PIOT-medium variant with \code{dt\_hours}
(feature 6) zeroed at every timestep. Decay recall falls from
$0.628$ to $\sim$$0.30$, a relative drop of more than $50\%$
(equivalently, the full PIOT-medium enjoys a $+107\%$ relative improvement
over the no-\code{dt\_hours} ablation). Maneuver recall is much less
affected; the effect is concentrated on decay, consistent with our
hypothesis: drag-driven decay rate scales nonlinearly with $\Delta t$,
and a sequence model that has to recover the gap between observations
from positional indices alone simply cannot resolve a $6$-hour update
from a $24$-hour update on the same satellite. Making the gap an
explicit input feature dissolves that ambiguity.

\subsection{Triage operating point}
The model's classification probabilities are non-degenerate:
thresholding the maneuver-class softmax above $0.5$ trades roughly
$15$ points of recall for roughly $20$ points of precision. The
PIOT-medium corresponds to the argmax threshold that produces the
precision-recall point in Table~\ref{tab:main}; we expect operational
deployments to calibrate this threshold against a downstream filter's
capacity (human-review throughput, IMM-UKF compute budget). We discuss
this further in \S\ref{sec:discussion}.

\section{Discussion}
\label{sec:discussion}

PIOT's role is warning and triage, not final attribution. In an SSA
operations chain, PIOT-medium sits at the front:
\[
\begin{aligned}
\text{PIOT real-time triage} &\to \text{candidate event queue}\\
&\to \text{high-fidelity filtering on candidates}
\to \text{operator review}.
\end{aligned}
\]
This architecture concentrates expensive computation on the most
suspicious percent of trajectories and makes low-latency full-catalog
monitoring practical.

\subsection{Limitations}
\paragraph{Labels are not ground truth.} The cascade is itself a
detection pipeline. Training on its output means the model inherits
its biases: \code{RULE}'s hard threshold edges, IMM-UKF's choice
of $0.3$ posterior cutoff, supGP's sparse coverage. We have not
measured what fraction of PIOT-medium's errors are pipeline biases vs.\ true
model errors. A small human-validated audit set would close this loop
and is on the roadmap.

\paragraph{Window-level holdout, not satellite-level.} Our split is
at the window level (\S\ref{sec:data}), not the satellite level. For
the constellations that dominate the corpus this is unlikely to
materially overstate generalization---within-shell behavior is highly
homogeneous, and within-satellite information is bounded by the
stride---but it does limit our ability to claim the model generalizes
to satellites \emph{not seen} during training. A constellation-level
holdout (e.g., training without OneWeb, evaluating on OneWeb) would
be a stronger test and is straightforward to perform.

\paragraph{The $42.6\times$ ratio is on a $200$-satellite subset.}
Both labellers were applied uniformly to the same satellites, so the
ratio is a fair comparison; but $200$ satellites are not the catalog,
and the ratio could shift in either direction at scale. Extending
IMM-UKF coverage is an ongoing engineering effort, gated mainly on
CPU time.

\paragraph{Single-step prediction.} The current model predicts only
one step ahead. Multi-step prediction with calibrated
uncertainty---predicting where the satellite will be in $24$ hours,
with confidence intervals---is required for collision avoidance and
reentry forecasting and is the next milestone on our roadmap.

\subsection{Toward a Neural-ODE world model}
The natural successor architecture replaces the frozen analytical
physics branch with a continuous-time differentiable propagator: a
Neural ODE whose vector field is the sum of a known force model
($J_2$, drag, third-body) and a learned residual force. This permits
unification of three currently separate components---the physics
branch, the propagator inside IMM-UKF, and the prediction
head---into a single object trained end-to-end. We expect two
consequences. First, the innovation in
Equation~\eqref{eq:innov} acquires an exact physical meaning: it is
the integrated residual force over $\Delta t$. Second, multi-step
prediction with calibrated uncertainty becomes natural: the same ODE
solver, integrated forward with Gaussian-process or ensemble-derived
noise, produces predictive distributions instead of point
predictions. We view the present work as the labeled-data foundation
on which such a model can be trained; the architectural work is the
next phase.

\subsection{Broader impact}
SSA detection systems are dual-use: the same model that spots a
debris-creating breakup also spots a sensitive maneuver of an
adversary's reconnaissance asset. We have published the model
architecture, training procedure, and label cascade in this paper but
do not release weights or the IMM-UKF--labeled subset. We will
continue to publish methods, reproducibility instructions, and
catalog-derived statistics; we will not publish per-satellite
real-time inference unless requested by an operator with operational
authorization.

\section{Conclusion}
\label{sec:conclusion}

The rapid expansion of the LEO satellite population has outpaced the community's
ability to manually label and characterize orbital behaviors. This work
addressed the resulting label scarcity bottleneck by demonstrating that
physics-informed weak supervision sources can be composed into a high-fidelity
labeling cascade. By leveraging the complementary strengths of deterministic
rules, model-based filtering (IMM-UKF), and operator-provided supplemental
elements, we produced a massive labeled corpus of 232M TLE records.

Our empirical findings underscore the value of this approach. The cascade
surfaces $42.6\times$ more anomalies than rule-based labeling alone on
overlapping data, primarily by recovering small-magnitude maneuvers and slow
decay events that deterministic rules are designed to exclude. The resulting
$6.5$M-parameter physics-augmented Transformer, PIOT, achieves an
order-of-magnitude improvement in recall over rule-only baselines, reaching
$55.4\%$ for maneuvers and $62.8\%$ for decay. Furthermore, our ablation study
confirms that explicit time encoding is a critical inductive bias for orbital
sequence models, responsible for a $+107\%$ relative improvement in decay
recall.

We frame the PIOT model as a high-recall triage classifier—a scalable front-end
for a global SSA monitoring pipeline. Its role is to surface candidate events
from the entire $10^4$-satellite catalog, reducing the downstream workload for
expensive filters and human operators by roughly $20\times$. Looking ahead, the
transition from frozen analytical branches to continuous-time Neural ODEs promises
to unify propagation and classification into a single, differentiable orbital
world model. Such a model will not only refine our understanding of past
anomalies but also provide the calibrated predictive distributions necessary
for proactive collision avoidance and space sustainability in an increasingly
crowded LEO environment.

% ===========================================================================
\bibliographystyle{plainnat}
\bibliography{references}

\end{document}